\advance\voffset by -20mm
\advance\hoffset by -15mm
\documentstyle[12pt]{article}
\begin{document}
\rightline{WIS-92/100/Dec-PH\quad}
\vspace{30mm}

\noindent{\Large{\bf NORMALIZATION OF SCATTERING STATES\\[2mm]
SCATTERING PHASE SHIFTS AND\\[2mm]
LEVINSON'S THEOREM}}
\vspace{17mm}

\noindent{\large Nathan Poliatzky$^*$
}\vspace{5mm}

\noindent Department of Physics, The Weizmann Institute of Science,
Rehovot, Israel
\vspace{28mm}

\noindent {\small {\em Abstract.}
We show that the normalization integral for the Schr\"odinger
and Dirac scattering wave functions contains, besides the usual
delta-function, a term proportional to the derivative of the
phase shift. This term is of zero measure with respect to the
integration over momentum variables and can be discarded in most cases.
Yet it carries the full information on phase shifts and can be used for
computation and manipulation of quantities which depend on phase shifts.
In this paper we prove Levinson's theorem in a most general way which
assumes only the completeness of states. In the case of a Dirac particle
we obtain a new result valid for positive and negative energies
separately. We also make a generalization of known results, for the
phase shifts in the asymptotic limit of high energies, to the case of
singular potentials. As an application we consider certain
equations, which arise in a generalized interaction picture of quantum
electrodynamics. Using the above mentioned results for the phase shifts
we prove that any solution of these equations, which has a finite number
of bound states, has a total charge zero. Furthermore, we show that in
these equations the coupling constant is not a free parameter, but rather
should be treated as an eigenvalue and hence must have a definite
numerical value.
}\vspace{17mm}

\section{INTRODUCTION AND RESULTS}\label{introduction}
It is a well known fact that the normalization integral of reduced
radial Schr\"odinger wave functions contains a delta-function. A less
well known fact is, that besides the delta-function, there are other
terms. In this paper we shall show that the precise expression is
\begin{eqnarray}&&\int_0^\infty{\rm d}r\;u_{kl}\left(r\right)u_{k'l}\left(r
\right)=2\pi\delta\left(k-k'\right)+\Delta\left(k,k'\right)\nonumber\\
&&\hspace{120pt}-\left(-1\right)^l\left\{2\pi\delta\left(k+k'\right)\cos\left[
\eta_l\left(k\right)+\eta_l\left(k'\right)\right]+\Delta\left(k,-k'\right)
\right\}\,,\label{norma}\end{eqnarray}
where
\begin{equation}\Delta\left(k,k'\right)=\left\{
\begin{array}{l}2\eta'_l\left(k\right)\,,\\0\,,\end{array}
\quad\begin{array}{l}k=k'\\k\neq k'\end{array}
\right.\,,\label{Deltakka}\end{equation}
$\eta'_l\left(k\right)=\partial_k\eta_l\left(k\right)$ and
$\eta_l\left(k\right)$ is a phase shift corresponding to the angular
momentum $l$ and linear momentum $k$. The corresponding result for the
case of Dirac equation is
\begin{equation}\int_0^\infty\hspace{-5pt}{\rm d}r\left[u_{1\epsilon\kappa}
\left(r\right)u_{1\epsilon'\kappa}\left(r\right)+u_{2\epsilon\kappa}\left(r
\right)u_{2\epsilon'\kappa}\left(r\right)\right]=\left\{\hspace{-5pt}
\begin{array}{l}2\pi\delta\left(k-k'\right)+\Delta\left(k,k'\right)\\
\quad-\left(-1\right)^l\left\{2\pi\delta\left(k+k'\right)\cos\left[
\eta_{\epsilon\kappa}\left(k\right)+\eta_{\epsilon\kappa}\left(k'\right)
\right]\right.\\
\qquad\left.+\Delta\left(k,-k'\right)\right\}\,,\quad\epsilon=\epsilon'\\
0\,,\quad\epsilon\neq \epsilon'\end{array}\right.\label{ndirintr}\end{equation}
where $\Delta\left(k,k'\right)$ is the same function as
in~(\ref{Deltakka}), except that the Schr\"odinger phase shift
$\eta_l$ is replaced by its relativistic counterpart
$\eta_{\epsilon\kappa}$. In most physical applications (for instance
the elastic scattering cross section) the normalization
integrals~(\ref{norma}) or~(\ref{ndirintr}) appear under the integral over
$k$ or $k'$, and hence the terms additional to the delta-function
$\delta\left(k-k'\right)$ do not contribute. Yet these terms contain
valuable information and can be used for the calculation of phase
shifts or for the manipulation of quantities depending on them. In
this paper we will use~(\ref{norma}) and~(\ref{ndirintr}) to prove
Levinson's theorem~\cite{Levinson} in a most direct and general way.

This theorem is one of the most interesting nonperturbative results in
quantum theory.
It has many potential applications and has been
applied recently in atomic physics~\cite{Spruchprl},\cite{Spruchpr},
in quantum field theories~\cite{Blank},\cite{Niemi}, and in solid state
physics (where it is known in a modified form under the name Friedel's
sum rule~\cite{Friedel}). In its original form the theorem says that
\begin{equation}\eta_l\left(0\right)=n_l\pi\,,\label{Levintr}\end{equation}
which relates the scattering phase shift $\eta_l\left(0\right)$ at
threshold (zero momentum) and for a given angular momentum $l$ to the
number of bound states $n_l$ of the Schr\"odinger equation with a
spherically symmetric potential. If the Schr\"odinger equation has a
zero-energy solution which vanishes at the origin and is finite at
infinity and yet not normalizable (it is called a half-bound
state or zero-energy resonance and is possible only if $l=0$) then, as
was first shown by R. Newton~\cite{Newtona}, Levinson's theorem is
modified to read
\begin{equation}\eta_0\left(0\right)=\left(n_0+\frac{1}{2}\right)\pi\,.
\label{LevNewt}\end{equation}
This result is subject to some restrictions on the potential. The most
general proof of Levinson's theorem was carried out by Ni \cite{Ni}
and Ma \cite{Ma} using the Sturm-Liouville theorem. Their result for
higher angular momenta $l\geq1$ is~(\ref{Levintr}), but for the S-state,
where a zero-energy resonance is possible, they obtained
\begin{equation}\eta_0\left(0\right)=n_0\pi+\frac{\pi}{2}\sin^2\eta_0\left(0
\right)\,.\label{Levmodintr}\end{equation}
The authors did not try to solve this equation for $\eta_0\left(0\right)$.
However, we point out that this equation is easily solved and there are
only three solutions:
\begin{equation}\eta_0\left(0\right)=\left\{\begin{array}{l}n_0\pi\\\left(n_0+
\frac{1}{4}\right)\pi\\\left(n_0+\frac{1}{2}\right)\pi\end{array}
\,.\right.\label{Levmodsstsolintr}\end{equation}
The first solution is valid for the case without a zero-energy resonance,
the other two if such a resonance exists. No examples are known for the
second solution. In this paper we will give yet a more general derivation
of the above results, which at the same time is considerably simpler.

The first correct statement of Levinson's theorem for Dirac particles
was given by Ma and Ni \cite{MaNi}
\begin{equation}\eta_{m\kappa}\left(0\right)+\eta_{-m\kappa}\left(0\right)=
\left(N_\kappa^++N_\kappa^-\right)\pi\,,\label{Levdintr}\end{equation}
which is valid whenever there is no threshold resonance and
\begin{equation}\eta_{m\kappa}\left(0\right)+\eta_{-m\kappa}\left(0\right)=
\left(N_\kappa^++N_\kappa^-\right)\pi+\left(-1\right)^l\frac{\pi}{2}\left[
\sin^2\eta_{m\kappa}\left(0\right)+\sin^2\eta_{-m\kappa}\left(0\right)\right]
\,,\label{Levmoddintr}\end{equation}
which is valid for the case with a threshold resonance (which can appear
only in the case $\kappa=\pm1$). Here $\pm m$ is the threshold energy
of the Dirac particle, $l=\vert\kappa\vert-1$ for $\kappa=-1,-2,
\ldots$ and $l=\kappa$ for $\kappa=1,2,\ldots$ is the orbital angular
momentum, $N_\kappa^+$ is the number of positive and $N_\kappa^-$ the
number of negative energy bound states of the Dirac equation and
$\eta_{\pm m\kappa}\left(0\right)$ are the phase shifts at threshold. Prior to
the work of Ma and Ni claims were published stating that
Levinson's theorem is valid for positive and negative energies separately
and in the same sense as in the nonrelativistic case, i.e.
$\eta_{\pm m\kappa}\left(0\right)=N_\kappa^\pm\pi$, but later such claims were
found incorrect \cite{MaNi}. However, we shall prove in this paper
that in a modified sense these claims are correct and that
\begin{equation}\hspace{6.4pt}\eta_{m\kappa}\left(0\right)=\left\{\begin{array}
{ll}n_l^+\pi,&l=0,1,\ldots\\\left(n_0^++\frac{1}{4}\right)\pi,&l=0\\\left(n_0^+
+\frac{1}{2}\right)\pi,\quad&l=0\end{array}\,,\right.
\label{Levwdsolpintr}\end{equation}
\begin{equation}\eta_{-m\kappa}\left(0\right)=\left\{\begin{array}{ll}
n_{\overline l}^-\pi,&\overline l=0,1,\ldots\\\left(n_0^-+\frac{1}{4}\right)\pi
,&\overline l=0\\\left(n_0^-+\frac{1}{2}\right)\pi,\quad&\overline l=0
\end{array}\,,\right.\label{Levwdsolmintr}\end{equation}
where $n_l^+$ and $n_{\overline l}^-$ are the numbers of bound state solutions
of certain Schr\"odinger equations which are given in the text and $\overline l
=l-\kappa/\vert\kappa\vert$. In~(\ref{Levwdsolpintr})
and~(\ref{Levwdsolmintr}) the first case refers to a situation without a
threshold resonance and the other two cases to a situation with a
threshold resonance. Equations~(\ref{Levwdsolpintr}) and~(\ref{Levwdsolmintr})
constitute the stronger statement of Levinson's theorem for Dirac
particles. As a consequence of~(\ref{Levdintr})
and~(\ref{Levwdsolpintr}),~(\ref{Levwdsolmintr}) it follows that
\begin{equation}N_\kappa^++N_\kappa^-=n_l^++n_{\overline l}^-\,,
\label{Nnpmintr}\end{equation}
whereas in general $N_\kappa^+\neq n_l^+$ and
$N_\kappa^-\neq n_{\overline l}^-$.

Besides the phase shifts at threshold one can also obtain
nonperturbatively the phase shifts in the asymptotic limit of high
energies. A well known result is (see ref.~\cite{Newton} p. 352)
\def\ar#1{\hbox to 0pt{\lower 5.0pt\hbox{\hskip4pt$\scriptstyle#1\rightarrow
            \infty$}\hss}\hbox to 30pt{\rightarrowfill}}
\begin{equation}\eta_l\left(k\right)\;\ar k\;-\frac{m}{k}\int_0^\infty{\rm d}r
\;V\left(r\right)\,,\label{etaklargeb}\end{equation}
where $V\left(r\right)$ is a spherically symmetric interaction potential,
and thus
\begin{equation}\eta_l\left(\infty\right)=0\,.\label{xx}\end{equation}
Obviously, equation~(\ref{etaklargeb}) is only valid if the integral on the
right-hand side exists, which is not the case for potentials with
a $1/r$ singularity or stronger. In this paper we shall extend
this result to the case where the potential at the origin may be as
singular as $1/r^{2-0}$ or less, including the obviously important case
$1/r$. At infinity the potential is assumed to vanish faster than $1/r$.
The result is
\begin{equation}\eta_l\left(k\right)\;\ar k\;k\int_0^\infty{\rm d}r\left[
\sqrt{1-\frac{2m}{k^2}V\left(r\right)}-1\right]\,,\label{xxx}\end{equation}
which still implies $\eta_l\left(\infty\right)=0$ and which reduces to
(\ref{etaklargeb}) for potentials less singular than $1/r$ at the origin.
In the case of a Dirac particle the Schr\"odinger result $\eta_l\left(\infty
\right)=0$ does not hold in general. Instead it was shown by
Parzen~\cite{Parzen} that
\begin{equation}\eta_{\epsilon\kappa}\left(k\right)\;\ar k\;-\frac{\epsilon}{k}
\int_0^\infty{\rm d}r\;V\left(r\right)=-\frac{\epsilon}{\vert\epsilon\vert}
\int_0^\infty{\rm d}r\;V\left(r\right)\,,\label{etakdlargeintr}\end{equation}
where $\epsilon=\pm\sqrt{k^2+m^2}$. In general, the right-hand side of
(\ref{etakdlargeintr}) is a constant which yields $\eta_{\pm\infty\kappa}\left(
\infty\right)=0$ only for a special type of potential. As was the case with
(\ref{etaklargeb}) the result~(\ref{etakdlargeintr}) only holds if the
integral on the right-hand side exists. This is not the case for
$V\left(r\right)\sim1/r$ at the origin. In this paper we shall
derive~(\ref{etakdlargeintr}) more rigorously than was done in
\cite{Parzen}, and we shall generalize the result to the case of a
potential behaving like $1/r$ at the origin. Unlike the Schr\"odinger
case there is no reason to treat more singular potentials since the
scattering wave functions do not behave well in this case. The result
we shall prove is
\begin{equation}\eta_{\epsilon\kappa}\left(k\right)\;\ar k\;k\int_0^\infty
{\rm d}r\;\left[\sqrt{1-\frac{2\epsilon}{k^2}V\left(r\right)}-1\right]\,,
\label{etakdlargeaintr}\end{equation}
which reduces to~(\ref{etakdlargeintr}) if $V$ is less singular than $1/r$ at
the origin.

As an application of the above results we shall investigate the following
set of equations
\begin{eqnarray}&&u'_{1\epsilon\kappa}+\frac{\kappa}{r}\,u_{1\epsilon\kappa}-
\left(m_0+\epsilon+\varphi\right)u_{2\epsilon\kappa}=0\nonumber\\
&&u'_{2\epsilon\kappa}-\frac{\kappa}{r}\,u_{2\epsilon\kappa}-\left(m_0-
\epsilon-\varphi\right)u_{1\epsilon\kappa}=0\,,
\label{radQEDdirintr}\end{eqnarray}
\begin{equation}\varphi''+\frac{2}{r}\varphi'=\frac{e_0^2}{4\pi}\;\frac{1}
{r^2}\sum_{\kappa=1}^\infty\kappa\left(\varrho_{\kappa}+\varrho_{-\kappa}
\right)\,,\label{radQEDmaxwintr}\end{equation}
\begin{eqnarray}&&\delta\left(r-r'\right)\delta_{ij}=\sum_{0\leq
\varepsilon_\kappa\leq m_0}u_{i\varepsilon_\kappa\kappa}\left(r\right)\;
u_{j\varepsilon_\kappa\kappa}\left(r'\right)+\sum_{0<\varepsilon_\kappa\leq
m_0}u_{i,-\varepsilon_\kappa\kappa}\left(r\right)\;u_{j,-\varepsilon_\kappa
\kappa}\left(r'\right)\nonumber\\ &&\hspace{75pt}+\int_{0+}^\infty
\frac{{\rm d}k}{2\pi}\left[u_{i\varepsilon\kappa}\left(r\right)u_{jE\kappa}
\left(r'\right)+u_{i,-\varepsilon\kappa}\left(r\right)u_{j,-\varepsilon\kappa}
\left(r'\right)\right]\,,\label{radQEDcomplintr}\end{eqnarray}
where $i=1,2$, $j=1,2$, $\kappa=\pm1,\pm2,\ldots\,$,
$u_{i,\pm\varepsilon_\kappa\kappa}$ are bound state and
$u_{i,\pm\varepsilon\kappa}$ are scattering state solutions
of~(\ref{radQEDdirintr}), $\epsilon=\pm\varepsilon_\kappa$,
$0\leq\varepsilon_\kappa\leq m_0$, are bound state and
$\epsilon=\pm\varepsilon$, $\varepsilon\equiv\sqrt{m_0^2+k^2}$, are
scattering state energies and
\begin{eqnarray}&&\varrho_\kappa=\sum_{0\leq \varepsilon_\kappa\leq m_0}\left(
u_{1\varepsilon_\kappa\kappa}^2+u_{2\varepsilon_\kappa\kappa}^2\right)-\sum_{
0<\varepsilon_\kappa\leq m_0}\left(u_{1,-\varepsilon_\kappa\kappa}^2+u_{2,
-\varepsilon_\kappa\kappa}^2\right)\nonumber\\ &&\hspace{25pt}+\int_{0+}^\infty
\frac{{\rm d}k}{2\pi}\left(u_{1\varepsilon\kappa}^2+u_{2\varepsilon\kappa}^2-
u_{1,-\varepsilon\kappa}^2 -u_{2,-\varepsilon\kappa}^2\right)\,.
\label{radQEDrhointr}\end{eqnarray}
For the sake of simplicity we assume that there are no threshold
resonances and hence in~(\ref{radQEDcomplintr}) and~(\ref{radQEDrhointr}) the
region of integration excludes $k=0$. These equations are subject to
certain boundary conditions which are explained in the text. Equation
(\ref{radQEDdirintr}) is the radial Dirac equation,~(\ref{radQEDmaxwintr}) is
one of the Maxwell equations (Poisson equation)
and~(\ref{radQEDcomplintr}) is the completeness relation for radial Dirac
wave functions. Equations (\ref{radQEDdirintr}-\ref{radQEDcomplintr}) arise as
a spherically symmetric special case of more general equations in a
recently proposed generalized interaction picture of quantum
electrodynamics (QED). The derivation of these equations from QED goes
beyond the framework of the present paper and we refer the reader to
the forthcoming paper~\cite{polQED}. In the present paper, using the
above results, we shall prove that for any solution of
(\ref{radQEDdirintr}-\ref{radQEDcomplintr}), which has a finite number of bound
states, the total charge vanishes:
\begin{equation}Q_0=0\,,\label{quantcharge}\end{equation}
where the charge density is defined by the right-hand side of
(\ref{radQEDmaxwintr}) (in units of $-e_0^2$). Furthermore, the coupling
constant $e_0^2/4\pi$ is not a free parameter but rather must have
a numerical value for which
\begin{equation}\int_0^\infty{\rm d}r\;\varphi\left(r\right)=0\,.
\label{QEDconsaintr}\end{equation}

\section{NORMALIZATION OF SCATTERING STATES}
\subsection{SCHR\"ODINGER CASE}\label{schroedinger}

Consider the radial Schr\"odinger equation
\begin{equation}u''_{kl}-\left[\frac{l\left(l+1\right)}{r^2}+2mV-k^2\right]
u_{kl}=0\,,\label{schreq}\end{equation}
for a scattering state characterized by the reduced radial wave
function $u_{kl}\left(r\right)$ subject to the boundary conditions
\begin{equation}u_{kl}\left(0\right)=0\,,\quad u_{kl}\left(r\right)\,\ar r\;2\,
\sin\left(kr-\frac{\pi l}{2}+\eta_l\left(k\right)\right)\,,
\label{bcond}\end{equation}
where $\eta_l\left(k\right)$ is the phase shift. Here we assume that the
potential $V\left(r\right)$ is less singular at the origin than
$1/r^2$ and that it vanishes at infinity faster than $1/r$. The
boundary conditions (\ref{bcond}) determine the normalization of the
wave functions:
\begin{eqnarray}&&\int_0^\infty{\rm d}r\;u_{kl}\left(r\right)u_{k'l}\left(r
\right)=2\pi\delta\left(k-k'\right)+\Delta\left(k,k'\right)\nonumber\\
&&\hspace{120pt}-\left(-1\right)^l\left\{2\pi\delta\left(k+k'\right)\cos\left[
\eta_l\left(k\right)+\eta_l\left(k'\right)\right]+\Delta\left(k,-k'\right)
\right\}\,,\qquad\label{norm}\end{eqnarray}
where
\begin{equation}\Delta\left(k,k'\right)=\left\{\begin{array}{ll}2\eta'_l\left(
k\right),\quad &k=k'\\0,&k\neq k'\end{array}\right.
\label{Deltakk}\end{equation}
and $\eta'_l\left(k\right)=\partial_k\eta_l\left(k\right)$. Notice
that in the physical region where both $k$ and $k'$ are positive the
last two terms in (\ref{norm}) vanish identically. Moreover, when
integrated over positive values of $k$ or $k'$, only the first
delta-function contributes. Thus in most cases one could drop all
terms except the first. Then, however, one has lost the valuable
information about the phase shifts $\eta_l$ contained in the
$\Delta\left(k,k'\right)$ term. Therefore it is especially desirable
to look into situations where only the diagonal terms in (\ref{norm})
are essential and the delta-functions do not contribute. For instance,
subtracting from~(\ref{norm}) its noninteracting counterpart the
delta-functions cancel and we obtain
\begin{equation}\int_0^\infty{\rm d}r\;\left[u_{kl}^2\left(r\right)-v_{kl}^2
\left(r\right)\right]=2\eta'_l\left(k\right)+\left(-1\right)^l2\pi\delta\left(
k\right)\sin^2\eta_l\left(k\right)\,,\qquad v_{kl}\left(r\right)=2kr j_l
\left(kr\right)\,,\label{diagn}\end{equation}
where $ j_l\left(kr\right)$  are the spherical Bessel functions.
This equation turns out to be quite useful, as will be shown below.
Also notice that the extra terms on the right-hand side
of~(\ref{norm}) cannot be cancelled by changing the normalization
in~(\ref{bcond}).

To derive equation~(\ref{norm}) we multiply~(\ref{schreq})
by $u_{k'l}$ and the corresponding equation for $u_{k'l}$ by
$u_{kl}$. Subtracting the resulting equations from one another
and integrating using $u_{kl}\left(0\right)=u_{k'l}\left(0\right)=0$, we obtain
\begin{equation}\int_0^R{\rm d}r\;u_{kl}\left(r\right)u_{k'l}\left(r\right)=
\frac{u_{kl}\left(R\right)u'_{k'l}\left(R\right)-u'_{kl}\left(R\right)u_{k'l}
\left(R\right)}{ k^2-k'^2}\,,\label{Rnorm}\end{equation}
where $u'_{kl}\left(R\right)=\partial_Ru_{kl}\left(R\right)$. For large enough
$R$ one can evaluate the right-hand side of~(\ref{Rnorm}) using the
asymptotic expression~(\ref{bcond}) and obtain
\begin{eqnarray}&&\hspace{-45pt}\frac{1}{2}\int_0^R{\rm d}r\,u_{kl}\left(r
\right)u_{k'l}\left(r\right)\nonumber\\&&\hspace{-25pt}=\frac{\sin\left[\left(
k-k'\right)R\right]}{k-k'}\cos\left[\eta_l\left(k\right)-\eta_l\left(k'\right)
\right]+\frac{\sin\left[\eta_l\left(k\right)-\eta_l\left(k'\right)\right]}
{k-k'}\cos\left[\left(k-k'\right)R\right]\nonumber\\&&\hspace{-24pt}-\left(-1
\right)^l\left\{\frac{\sin\left[\left(k+k'\right)R\right]}{k+k'}\cos\left[
\eta_l\left(k\right)+\eta_l\left(k'\right)\right]+\frac{\sin\left[\eta_l\left(
k\right)+\eta_l\left(k'\right)\right]}{k+k'}\cos\left[\left(k+k'\right)R\right]
\right\}.\label{Rnorma}\end{eqnarray}
Using $\eta_l\left(-k\right)=-\eta_l\left(k\right)$ (see ref.~\cite{Baz} for
a proof), equation~(\ref{norm}) follows as $R\rightarrow\infty$.\par
It is instructive to consider a somewhat different derivation. Taking the
derivative of~(\ref{schreq}) with respect to $k$ and multiplying the result by
$u_{kl}$, then multiplying~(\ref{schreq}) by $\partial_ku_{kl}$,
subtracting the resulting two equations from one another and
integrating, we obtain
\begin{equation}\int_0^R{\rm d}r\;u_{kl}^2\left(r\right)=\frac{1}{2k}\left[
u'_{kl}\left(R\right)\partial_ku_{kl}\left(R\right)-u_{kl}\left(R\right)
\partial_ku'_{kl}\left(R\right)\right]\,.\label{Rnormb}\end{equation}
For a large enough $R$ we can use the asymptotics~(\ref{bcond}) to evaluate
the right-hand side of~(\ref{Rnormb}). The result
\begin{equation}\int_0^R{\rm
d}r\;u_{kl}^2\left(r\right)=2R+2\eta'_l\left(k\right)-\left(-1\right)^l\frac{1}
{ k}\,\sin\left[2kR+2\eta_l\left(k\right)\right)\label{Rnormc}\end{equation}
was first obtained by
L\"uders \cite{Lued}. Equation~(\ref{Rnormc}) is exact if the
potential $V\left(r\right)$ vanishes for $r>R$ and is valid asymptotically
otherwise. From~(\ref{Rnormc}) one easily recovers the diagonal terms of
(\ref{norm}).

\subsection{DIRAC CASE}

Consider now the reduced radial Dirac equations
\begin{eqnarray}&&u'_{1\epsilon\kappa}+\frac{\kappa}{r}\,u_{1\epsilon\kappa}-
\left(\epsilon+m-V\right)u_{2\epsilon\kappa}=0\nonumber\\&&u'_{2\epsilon\kappa}
-\frac{\kappa}{r}\,u_{2\epsilon\kappa}+\left(\epsilon-m-V\right)u_{1\epsilon
\kappa}=0\,,\label{raddir}\end{eqnarray}
where $\epsilon$ is the energy, $V\left(r\right)$ is the time component of a
vector potential and $\kappa=\pm1,\pm2,\ldots\,$. The quantum number
$\kappa$ is the standard parametrization of the total angular momentum
$j=\vert\kappa\vert-1/2$ and of the relative orientation between the
spin and the orbital angular momentum. The appropriate boundary
conditions are
\begin{eqnarray}&&u_{1\epsilon\kappa}\left(0\right)=0\,,\quad u_{1\epsilon
\kappa}\left(r\right)\;\ar r\;\,\sqrt{\frac{\epsilon+m}{2\epsilon}}\,2\,
\sin\left(kr-\frac{\pi l}{2}+\eta_{\epsilon\kappa}\left(k\right)\right)\,,
\label{bcondo}\\&&u_{2\epsilon\kappa}\left(0\right)=0\,,\quad u_{2\epsilon
\kappa}\left(r\right)\;\ar r\;\,\sqrt{\frac{\epsilon+m}{2\epsilon}}\,
\frac{\kappa_0k}{\epsilon+m}\,2\,\sin\left(kr-\frac{\pi\overline l}{2}+
\eta_{\epsilon\kappa}\left(k\right)\right)\,,\label{bcondt}\end{eqnarray}
where $k=\pm\sqrt{\epsilon^2-m^2}$, $\kappa_0=\kappa/\vert\kappa\vert$,
$l=\vert\kappa\vert-\left(1-\kappa_0\right)/2$, $\overline l=l-\kappa_0$, and
$\eta_{\epsilon\kappa}\left(k\right)$ is the phase shift. To ensure
the consistency of~(\ref{bcondo}) and~(\ref{bcondt})
with~(\ref{raddir}) we assume that $V\left(r\right)$ behaves like or
less singularly than $1/r$ at the origin and that it vanishes at
infinity faster than $1/r$. Thus the Yukawa potential
${\rm e}^{-\mu_0r}/r$ is allowed but the Coulomb potential is excluded.
The normalization of the wave functions resulting from the boundary
conditions~(\ref{bcondo}) and~(\ref{bcondt}) is
\begin{equation}\int_0^\infty\hspace{-5pt}{\rm d}r\left[u_{1\epsilon\kappa}
\left(r\right)u_{1\epsilon'\kappa}\left(r\right)+u_{2\epsilon\kappa}\left(r
\right)u_{2\epsilon'\kappa}\left(r\right)\right]=\left\{\hspace{-5pt}
\begin{array}{l}2\pi\delta\left(k-k'\right)+\Delta\left(k,k'\right)\\
\quad-\left(-1\right)^l\left\{2\pi\delta\left(k+k'\right)\cos\left[
\eta_{\epsilon\kappa}\left(k\right)+\eta_{\epsilon\kappa}\left(k'\right)
\right]\right.\\
\qquad\left.+\Delta\left(k,-k'\right)\right\}\,,\quad\epsilon=\epsilon'\\
0\,,\quad\epsilon\neq \epsilon'\end{array}\right.\label{ndir}\end{equation}
where $\Delta\left(k,k'\right)$ is the same function as
in~(\ref{Deltakk}), except that the phase shift $\eta_l$ is replaced
by $\eta_{\epsilon\kappa}$. Notice the similarity of~(\ref{ndir}) to
the corresponding Schr\"odinger case~(\ref{norm}). As in that case we
can extract the information on the phase shifts
$\eta_{\epsilon\kappa}$ by subtracting from~(\ref{ndir}) the
corresponding equation for the noninteracting case, and we obtain
\begin{equation}\int_0^\infty{\rm d}r\,\left[u_{1\epsilon\kappa}^2\left(r
\right)+u_{2\epsilon\kappa}^2\left(r\right)-v_{1\epsilon\kappa}^2\left(r
\right)-v_{2\epsilon\kappa}^2\left(r\right)\right]=2\eta'_{\epsilon\kappa}
\left(k\right)+\left(-1\right)^l2\pi\delta\left(k\right)\sin^2\eta_{\epsilon
\kappa}\left(k\right)\,,\label{diagndir}\end{equation}
where
\begin{equation}v_{1\epsilon\kappa}\left(r\right)=\sqrt{\frac{\epsilon+m}
{2\epsilon}}\,2kr\, j_l\left(kr\right)\,,\quad v_{2\epsilon\kappa}\left(r
\right)=\sqrt{\frac{\epsilon+m}{2\epsilon}}\,\frac{\kappa_0k}{\epsilon+m}\,
2kr\, j_{\overline l}\left(kr\right)\label{freewfintr}\end{equation}
is the regular free-particle solution of the Dirac equation~(\ref{raddir}),
which obeys~(\ref{bcondo}),~(\ref{bcondt}) and~(\ref{ndir}) with
$\eta_{\epsilon\kappa} \left(k\right)=0$. Contrary to the case of the
Schr\"odinger equation the energy can now be either positive or
negative and we can use $u_{i,-\epsilon \kappa}$ instead of
$v_{i\epsilon\kappa}$ to extract the diagonal terms in~(\ref{ndir}).
The result is
\begin{eqnarray}&&\hspace{-40pt}\int_0^\infty{\rm d}r\,\left[u_{1\epsilon
\kappa}^2\left(r\right)+u_{2\epsilon\kappa}^2\left(r\right)-u_{1,-\epsilon
\kappa}^2\left(r\right)-u_{2,-\epsilon\kappa}^2\left(r\right)\right]\nonumber
\\&&\hspace{-23pt}=2\left[\eta'_{\epsilon\kappa}\left(k\right)-\eta'_{-\epsilon
\kappa}\left(k\right)\right]+\left(-1\right)^l2\pi\delta\left(k\right)\sin
\left[\eta_{\epsilon\kappa}\left(k\right)+\eta_{-\epsilon\kappa}\left(k\right)
\right]\sin\left[\eta_{\epsilon\kappa}\left(k\right)-\eta_{-\epsilon\kappa}
\left(k\right)\right].\label{diagndira}\end{eqnarray}

To derive~(\ref{ndir}) we multiply the Dirac equation~(\ref{raddir}) by
$\left(-u_{2\epsilon'\kappa},u_{1\epsilon'\kappa}\right)$ and the Dirac
equation for $\epsilon'$ by
$\left(-u_{2\epsilon\kappa},u_{1\epsilon\kappa}\right)$ and subtract the
results. Upon integration we obtain
\begin{equation}\int_0^R{\rm d}r\,\left[u_{1\epsilon\kappa}\left(r\right)
u_{1\epsilon'\kappa}\left(r\right)+u_{2\epsilon\kappa}\left(r\right)
u_{2\epsilon'\kappa}\left(r\right)\right]=\frac{u_{1\epsilon\kappa}\left(R
\right)u_{2\epsilon'\kappa}\left(R\right)-u_{1\epsilon'\kappa}\left(R\right)
u_{2\epsilon\kappa}\left(R\right)}{\epsilon-\epsilon'}\,.
\label{ndira}\end{equation}
For a large enough $R$ the right-hand side of~(\ref{ndira}) can be
evaluated using the asymptotic wave functions~(\ref{bcondo})
and~(\ref{bcondt}). The result is
\begin{eqnarray}&&\hspace{-20pt}\int_0^R{\rm d}r\,\left[u_{1\epsilon\kappa}
\left(r\right)u_{1\epsilon'\kappa}\left(r\right)+u_{2\epsilon\kappa}\left(r
\right)u_{2\epsilon'\kappa}\left(r\right)\right]\label{ndirb}\\&&\hspace{-12pt}
=A\left\{\frac{\sin\left[\left(k-k'\right)R\right]}{k-k'}\cos\left[
\eta_{\epsilon\kappa}\left(k\right)-\eta_{\epsilon'\kappa}\left(k'\right)
\right]+\frac{\sin\left[\eta_{\epsilon\kappa}\left(k\right)-\eta_{\epsilon'
\kappa}\left(k'\right)\right]}{k-k'}\cos\left[\left(k-k'\right)R\right]\right\}
\nonumber\\&&\hspace{-12pt}-\left(-1\right)^lB\left\{\frac{\sin\left[\left(k+k'
\right)R\right]}{k+k'}\cos\left[\eta_{\epsilon\kappa}\left(k\right)+
\eta_{\epsilon'\kappa}\left(k'\right)\right]+\frac{\sin\left[\eta_{\epsilon
\kappa}\left(k\right)+\eta_{\epsilon'\kappa}\left(k'\right)\right]}{k+k'}
\cos\left[\left(k+k'\right)R\right]\right\}.\nonumber\end{eqnarray}
where
\begin{eqnarray}A=\left(\frac{\epsilon}{\vert\epsilon\vert}k\sqrt{
\frac{\epsilon'+m}{\epsilon\epsilon'\left(\epsilon+m\right)}}+\frac{\epsilon'}
{\vert\epsilon'\vert}k'\sqrt{\frac{\epsilon+m}{\epsilon\epsilon'\left(
\epsilon'+m\right)}}\;\right)\frac{\epsilon+\epsilon'}{k+k'}\,,
\hspace{100pt}&&\nonumber\\B=\left(\frac{\epsilon}{\vert\epsilon\vert}k\sqrt{
\frac{\epsilon'+m}{\epsilon\epsilon'\left(\epsilon+m\right)}}-\frac{\epsilon'}
{\vert\epsilon'\vert}k'\sqrt{\frac{\epsilon+m}{\epsilon\epsilon'\left(
\epsilon'+m\right)}}\;\right)\frac{\epsilon+\epsilon'}{k-k'}\,.&&
\label{ndirwhere}\end{eqnarray}
Equation~(\ref{ndir}) is now easily recovered as $R\rightarrow\infty$.

As in the Schr\"odinger case the diagonal terms of~(\ref{ndir}) can be
obtained somewhat differently. We derive the Dirac equation
(\ref{raddir}) with respect to $\epsilon$, multiply the resulting equation
by $\left(-u_{2\epsilon\kappa},u_{1\epsilon\kappa}\right)$ and
subtract from the result the Dirac equation multiplied by
$\left(-\partial_\epsilon u_{2\epsilon\kappa},\partial_\epsilon u_{1\epsilon
\kappa}\right)$. Upon integration we obtain
\begin{equation}\int_0^R{\rm d}r\,\left[u_{1\epsilon\kappa}^2\left(r\right)
+u_{2\epsilon\kappa}^2\left(r\right)\right]=u_{2\epsilon\kappa}\left(R\right)
\partial_\epsilon u_{1\epsilon\kappa}\left(R\right)-u_{1\epsilon\kappa}\left(
R\right)\partial_\epsilon u_{2\epsilon\kappa}\left(R\right)\,.
\label{ndirc}\end{equation}
Taking $R$ large and inserting the asymptotic wave functions~(\ref{bcondo})
and~(\ref{bcondt}) on the right-hand side, we obtain
\begin{equation}\int_0^R{\rm d}r\,\left[u_{1\epsilon\kappa}^2\left(r\right)
+u_{2\epsilon\kappa}^2\left(r\right)\right]=2R+2\eta'_{\epsilon\kappa}\left(
k\right)-\left(-1\right)^l\frac{m}{\epsilon k}\,\sin\left[2kR+2\eta_{\epsilon
\kappa}\left(k\right)\right]\label{Rnormdir}\end{equation}
{}From~(\ref{Rnormdir}) one easily recovers the diagonal terms of~(\ref{ndir}).
Notice the close similarity of~(\ref{Rnormdir}) to the corresponding
result for the Schr\"odinger case~(\ref{Rnormc}). In particular for small
energies ($m/\epsilon\rightarrow1$) both results coincide.

\section{HIGH ENERGY LIMIT}

\subsection{SCHR\"ODINGER CASE}

It is a well known result (see ref.~\cite{Newton} p. 352) that
\begin{equation}\eta_l\left(k\right)\;\ar k\;-\frac{m}{k}\int_0^\infty{\rm d}
r\;V\left(r\right)\label{etaklarge}\end{equation}
and thus
\begin{equation}\eta_l\left(\infty\right)=0\,.\label{etainf}\end{equation}
Obviously, equation~(\ref{etaklarge}) is only valid if the integral on the
right-hand side exists, which is not the case for potentials with
a $1/r$ singularity or stronger. In this section we want to extend
this result to the case where the potential at the origin may be as
singular as $1/r^{2-0}$ or less, including the obviously important case
$1/r$. At infinity the potential is assumed to vanish faster than $1/r$.
The result we shall now prove is
\begin{equation}\eta_l\left(k\right)\;\ar k\;k\int_0^\infty{\rm d}r\;\left[
\sqrt{1-\frac{2m}{ k^2}V\left(r\right)}-1\right]\,,
\label{etaklargea}\end{equation}
which still implies~(\ref{etainf}) and which reduces to~(\ref{etaklarge}) for
potentials less singular than $1/r$ at the origin.

We introduce a new variable $\overline r$ and a new wave function
$\chi_{kl}\left(\overline r\right)$
\begin{equation}u_{kl}\left(r\right)=\frac{\chi_{kl}\left(\overline r\right)}
{\sqrt{F}}\,,\qquad\overline r=\int_0^r{\rm d}r'\;F\left(r'\right)\,,\qquad
F\left(r\right)=\sqrt{1-\frac{2m}{ k^2}V\left(r\right)}\,.
\label{rbar}\end{equation}
Putting this into the Schr\"odinger equation~(\ref{schreq}) we get
\begin{equation}\ddot{\chi}_{kl}-\left[\frac{l\left(l+1\right)}{\overline r^2}
\left(1+\frac{1}{k}A\right)-\frac{m}{2k^2}B-\frac{3m}{4k^4}C-k^2\right]
\chi_{kl}=0\,,\label{chieq}\end{equation}
where
\begin{equation}A=\frac{k\frac{\overline r^2-r^2}{ r^2}+\frac{2m}{k}V}
{1-\frac{2m}{ k^2}V}\,,\quad B=\frac{\ddot{V}}{1-\frac{2m}{k^2}V}\,,\quad
C=\left(\frac{\dot{V}}{1-\frac{2m}{ k^2}V}\right)^2\label{ABCchi}\end{equation}
and where the dots mean the derivatives with respect to the new variable
$\overline r$. Consider now a sphere of radius $\overline
r=\sqrt{m/k}$ around the origin, $k$ being large, and assume the case
of the strongest allowed singularity $V\sim1/r^{2-0}$. Then everywhere
outside of the sphere the terms proportional to $A$, $B$ and $C$
in~(\ref{chieq}) are bounded by certain positive powers of $m/k$.
Hence as $k\rightarrow\infty$, we obtain
\begin{equation}\ddot{\chi}_{kl}-\left[\frac{l\left(l+1\right)}{\overline r^2}
-k^2\right]\chi_{kl}=0\,,\label{chieqa}\end{equation}
of which the physical solution is $\chi_{kl}\left(\overline r\right)=2k
\overline r j_l\left(k\overline r\right)$. Therefore as
$k\rightarrow\infty$, from~(\ref{rbar}) we have
\begin{equation}u_{kl}\left(r\right)\;\ar r\;2\,\sin\left(kr-\frac{\pi l}{2}+
k\int_0^\infty{\rm d}r\;\left[\sqrt{1-\frac{2m}{k^2}V\left(r\right)}-1\right]
\right)\label{etainfa}\end{equation}
and hence~(\ref{etaklargea}).

\subsection{DIRAC CASE}

In the case of a Dirac particle the Schr\"odinger result~(\ref{etainf})
in general does not hold. Instead it was shown by Parzen \cite{Parzen} that
\begin{equation}\eta_{\epsilon\kappa}\left(k\right)\;\ar k\;-\frac{\epsilon}{k}
\int_0^\infty{\rm d}r\;V\left(r\right)=-\frac{\epsilon}{\vert\epsilon\vert}
\int_0^\infty{\rm d}r\;V\left(r\right)\,,\label{etakdlarge}\end{equation}
where $\epsilon=\pm\sqrt{k^2+m^2}$. In general, the right-hand side of
(\ref{etakdlarge}) is a constant which yields $\eta_{\pm\infty\kappa}\left(
\infty\right)=0$  only for a special type of potential. As was the case with
(\ref{etaklarge}), the result~(\ref{etakdlarge}) only holds if the integral on
the right-hand side exists. This is not the case for $V\left(r\right)\sim1/r$
at the origin. In this section we want to rederive~(\ref{etakdlarge}) in a more
rigorous way than was done in \cite{Parzen} and to extend the result to the
case of a potential behaving like $1/r$ at the origin. Unlike the
Schr\"odinger case there is no reason to treat more singular
potentials since the scattering wave functions do not behave well in
this case. The result we shall prove is
\begin{equation}\eta_{\epsilon\kappa}\left(k\right)\;\ar k\;k\int_0^\infty{\rm
d}r\;\left[\sqrt{1-\frac{2\epsilon}{ k^2}V\left(r\right)}-1\right]\,,
\label{etakdlargea}\end{equation}
which reduces to~(\ref{etakdlarge}) if $V$ is less singular than $1/r$ at
the origin.

Eliminating $u_{2\epsilon\kappa}$ from~(\ref{raddir}), we obtain
\begin{equation}u_{1\epsilon\kappa}''-\left[\frac{\kappa\left(\kappa+1\right)}
{r^2}-\frac{\kappa}{ r}\frac{V'}{\epsilon+m-V}-V^2+2\epsilon V-k^2\right]
u_{1\epsilon\kappa}+\frac{V'}{\epsilon+m-V}\,u_{1\epsilon\kappa}'=0\,.
\label{qdireqo}\end{equation}
Eliminating $u_{1\epsilon\kappa}$ from (\ref{raddir}) leads to
\begin{equation}u_{2\epsilon\kappa}''-\left[\frac{\kappa\left(\kappa-1\right)}
{r^2}+\frac{\kappa}{ r}\frac{V'}{\epsilon-m-V}-V^2+2\epsilon V-k^2\right]
u_{2\epsilon\kappa}+\frac{V'}{\epsilon-m-V}\,u_{2\epsilon\kappa}'=0\,.
\label{qdireqt}\end{equation}
Equations~(\ref{qdireqo}) and~(\ref{qdireqt}) are equivalent to the
Dirac equation (\ref{raddir}) provided the boundary
conditions~(\ref{bcondo}) and~(\ref{bcondt}) are imposed. Actually it
is not necessary to solve~(\ref{qdireqo}) and (\ref{qdireqt})
simultaneously. For instance, if~(\ref{qdireqo}) is solved for
$u_{1\epsilon\kappa}$ one gets $u_{2\epsilon\kappa}$ through the first
of equations (\ref{raddir}). In order to get rid of the last term
in~(\ref{qdireqo}) and (\ref{qdireqt}) we introduce the new wave
functions $\left( w_{1\epsilon\kappa},w_{2\epsilon\kappa}\right)$
defined through
\begin{equation}u_{1\epsilon\kappa}\left(r\right)=\sqrt{\frac{\epsilon+m-V
\left(r\right)}{2\epsilon}}\,w_{1\epsilon\kappa}\left(r\right)\,,\quad u_{2
\epsilon\kappa}\left(r\right)=\frac{\epsilon k\kappa}{\vert\epsilon\vert\vert
k\vert\vert\kappa\vert}\sqrt{\frac{\epsilon-m-V\left(r\right)}{2\epsilon}}\,
w_{2\epsilon\kappa}\left(r\right)\,.\label{wdir}\end{equation}
Putting these in~(\ref{qdireqo}) and~(\ref{qdireqt}), we obtain
\begin{eqnarray}w_{1\epsilon\kappa}''-\left[\frac{l\left(l+1\right)}{r^2}-
\frac{\kappa}{r}\frac{V'}{\epsilon+m-V}+\frac{1}{2}\frac{V''}{\epsilon+m-V}
\right.\hspace{167pt}&&\nonumber\\\left.+\frac{3}{4}\left(\frac{V'}{\epsilon+
m-V}\right)^2-V^2+2\epsilon V-k^2\right]w_{1\epsilon\kappa}=0\,,&&
\label{wdireqo}\end{eqnarray}
\begin{eqnarray}w_{2\epsilon\kappa}''-\left[\frac{\overline l\left(\overline l
+1\right)}{r^2}+\frac{\kappa}{r}\frac{V'}{\epsilon-m-V}+\frac{1}{2}\frac{V''}
{\epsilon-m-V}\right.\hspace{167pt}&&\nonumber\\\left.+\frac{3}{4}\left(
\frac{V'}{\epsilon-m-V}\right)^2-V^2+2\epsilon V-k^2\right]w_{2\epsilon\kappa}
=0\,,&&\label{wdireqt}\end{eqnarray}
where we used $\kappa\left(\kappa+1\right)=l\left(l+1\right)$ and
$\kappa\left(\kappa-1\right)=\overline l\left(\overline l+1\right)$
which are readily obtained from $l=\vert\kappa\vert-\left(1-\kappa_0\right)/2$
and $\overline l=l-\kappa_0$, $\kappa_0=\kappa/\vert\kappa\vert$.
Equations~(\ref{wdireqo}) and~(\ref{wdireqt}) are of the
Schr\"odinger-type except that the potential depends on the energy.
Solving~(\ref{wdireqo}) and~(\ref{wdireqt}) is equivalent to solving
the original Dirac equation~(\ref{raddir}) provided the boundary conditions
\begin{equation} w_{1\epsilon\kappa}\left(0\right)=0\,,\quad w_{1\epsilon
\kappa}\left(r\right)\;\ar r\;2\,\sin\left(kr-\frac{\pi l}{2}+\eta_{\epsilon
\kappa}\left(k\right)\right)\,,\label{wbcondo}\end{equation}
\begin{equation} w_{2\epsilon\kappa}\left(0\right)=0\,,\quad w_{2\epsilon
\kappa}\left(r\right)\;\ar r\;2\,\sin\left(kr-\frac{\pi\overline l}{2}+
\eta_{\epsilon\kappa}\left(k\right)\right)\label{wbcondt}\end{equation}
are met. We now turn to the limit $k\rightarrow\infty$. To perform this limit
we can use either~(\ref{wdireqo}) or~(\ref{wdireqt}), the result will
be the same. As in the Schr\"odinger case, considering a small sphere
of radius $\left(m/k\right)^{1/4}$, we observe that everywhere outside
of the sphere the second, third and fourth terms in the parentheses
of~(\ref{wdireqo}) and (\ref{wdireqt}) are bounded by positive powers
of $m/k$ as $k\rightarrow\infty$ and hence can be neglected. The term
$V^2$ behaves as $1/r^2$ at the origin and vanishes faster than
$1/r^2$ at infinity. Hence it does alter the form of the wave function
at the origin, but at large distances it contributes only to the phase
of the wave function, which remains a Bessel function. However, $V^2$
is of order zero in $k$ so that, according to the preceding section,
it does not contribute to the asymptotic limit
$\eta_{\epsilon\kappa}\left(\infty\right)$ and, in particular, cannot
compete with the term $2\epsilon V$ which is of order one in $k$.
Hence as $k\rightarrow\infty$ we neglect it and obtain from~(\ref{wdireqo})
\begin{equation}w_{1\epsilon\kappa}''-\left[\frac{l\left(l+1\right)}{r^2}+2
\epsilon V-k^2\right]w_{1\epsilon\kappa}=0,\label{wdireqoklarge}\end{equation}
which is subject to the boundary condition~(\ref{wbcondo}). Thus the high
energy limit of the Dirac equation is dominated by the Schr\"odinger
equation~(\ref{schreq}) in which the mass $m$ is replaced by the energy
$\epsilon=\pm\sqrt{k^2+m^2}$. It is easy to check that, despite
this alteration of the Schr\"odinger equation, the arguments of the
preceding section are still valid and we can apply~(\ref{etaklargea})
replacing $m$ by $\epsilon$. The result is given in~(\ref{etakdlargea}).

\section{\hspace{-9.7pt}ZERO ENERGY LIMIT: LEVINSON'S THEOREM}

\subsection{SCHR\"ODINGER CASE}

The set of all physical solutions of the Schr\"odinger equation~(\ref{schreq})
constitutes a complete set in the sense that for each $l=0,1,\ldots$
\begin{equation}\sum_{\epsilon_l<0}u_{\epsilon_ll}\left(r\right)u_{\epsilon_ll}
\left(r'\right)+\int_0^\infty\frac{{\rm d}k}{2\pi}\,u_{kl}\left(r\right)u_{kl}
\left(r'\right)=\delta\left(r-r'\right)\,,\label{compl}\end{equation}
where $u_{\epsilon_ll}$ are bound state solutions of~(\ref{schreq})
normalized according to
\begin{equation}\int_0^\infty{\rm d}r\;u_{\epsilon_ll}^2\left(r\right)=1
\label{bstnorm}\end{equation}
and $u_{kl}$ are scattering state solutions~(\ref{bcond}). With respect to
the completeness relation we have to distinguish two cases. In the first
case, if there exists a zero-energy solution of the Schr\"odinger
equation~(\ref{schreq}), which vanishes at the origin and is finite at large
distances (we will call it finite), it is not normalizable in the
sense of~(\ref{bstnorm}). Such a solution, clearly, is not a bound state
(it is called a half-bound state \cite{Newton} or a
zero-energy resonance \cite{Taylor} since it becomes a
bound state after an arbitrarily small increase in the strength of the
potential). It is part of the continuum and is responsible for the
delta-function in~(\ref{diagn}). Hence the sum over bound states on the
left-hand side of~(\ref{compl}) does not include the zero-energy solution
and the integral over $k$ is understood as including the value $k=0$.
This is the meaning of the completeness relation~(\ref{compl}). In the second
case, if a finite, zero-energy, solution exists, it is normalizable. Such
a solution is a bound state and must be included in the sum over
bound states on the left-hand side of~(\ref{compl}) and, consequently, the
integral over $k$ is understood as excluding the value $k=0$. The
completeness relation for this case is
\begin{equation}\sum_{\epsilon_l\leq0}u_{\epsilon_ll}\left(r\right)
u_{\epsilon_ll}\left(r'\right)+\int_{0+}^\infty\frac{{\rm d}k}{2\pi}\,u_{kl}
\left(r\right)u_{kl}\left(r'\right)=\delta\left(r-r'\right)\,.
\label{complplus}\end{equation}

Notice that, if no finite zero-energy solution exists, then one can use
either one of the completeness relations. In case a finite zero-energy
solution exists, it is necessarily a zero-energy resonance if $l=0$,
and a bound state if $l\ge1$. This is easily seen by examining the
Schr\"odinger equation~(\ref{schreq}) at large distances for potentials
which vanish faster than $1/r^2$, and noticing that a finite zero-energy
solution behaves as $1/r^l$. If the potential does not vanish faster than
$1/r^2$ at infinity, then finite, zero-energy, solutions do not exist.

Obviously the completeness relation also holds for a free particle,
except that in this case there are no bound states and hence
(\ref{compl}) and~(\ref{complplus}) are identical. We now proceed with
the second case above using~(\ref{complplus}). Subtracting
from~(\ref{complplus}) the corresponding equation for the
noninteracting case the delta-functions cancel and we obtain
\begin{equation}\sum_{\epsilon_l\leq0}u_{\epsilon_ll}\left(r\right)
u_{\epsilon_ll}\left(r'\right)+\int_{0+}^\infty\frac{{\rm d}k}{2\pi}\,\left[
u_{kl}\left(r\right)u_{kl}\left(r'\right)- v_{kl}\left(r\right)v_{kl}\left(
r'\right)\right]=0\,,\label{compla}\end{equation}
where $ v_{kl}$ are the free particle solutions. Thus the diagonal
part of~(\ref{compla}) reads
\begin{equation}\sum_{\epsilon_l\leq0}u_{\epsilon_ll}^2\left(r\right)+
\int_{0+}^\infty\frac{{\rm d}k}{2\pi}\,\left[u_{kl}^2\left(r\right)-v_{kl}^2
\left(r\right)\right]=0\,.\label{complb}\end{equation}
Integrating over $r$ and substituting~(\ref{diagn}), we obtain
\begin{equation}\eta_l\left(0\right)=\eta_l\left(\infty\right)+n_l\pi
\label{Levinf}\end{equation}
where $n_l$ is the number of bound states having angular momentum $l$ (not
counting the $2l+1$ degeneracy). Finally substituting~(\ref{etainf}), we get
\begin{equation}\eta_l\left(0\right)=n_l\pi\,.\label{Lev}\end{equation}
Equation~(\ref{Lev}) is known as Levinson's theorem \cite{Levinson}.
It is valid whenever there is no zero-energy resonance and
this, as explained above, is the case for all $l\ge1$. This is, of
course, a well known result.

We now return to the first case where a finite zero-energy solution,
if it exists, is not normalizable and where we have to use~(\ref{compl})
instead of~(\ref{complplus}). Since a zero-energy resonance can occur only
for $l=0$ we do not need to consider higher angular momentum states.
Because the region of integration in~(\ref{compl}) contains the value $k=0$,
the steps preceding~(\ref{Lev}) remain the same except that now the
delta-function on the right-hand side of~(\ref{diagn}) contributes, and we
obtain
\begin{equation}\eta_0\left(0\right)=\eta_0\left(\infty\right)+n_0\pi+
\frac{\pi}{2}\sin^2\eta_0\left(0\right)\,.\label{Levmodinf}\end{equation}
Substituting~(\ref{etainf}), we get
\begin{equation}\eta_0\left(0\right)=n_0\pi+\frac{\pi}{2}\sin^2\eta_0\left(0
\right)\,.\label{Levmod}\end{equation}
This version of Levinson's theorem was first derived by Ni \cite{Ni}.
Equation~(\ref{Levmod}) can be solved for $\eta_0\left(0\right)$ and
there are only three solutions:
\begin{equation}\eta_0\left(0\right)=\left\{\begin{array}{l}n_0\pi\\\left(n_0+
\frac{1}{4}\right)\pi\\\left(n_0+\frac{1}{2}\right)\pi\end{array}
\,.\right.\label{Levmodsstsol}\end{equation}
The second and third solutions correspond to the case with a
zero-energy resonance, whereas the first solution corresponds to the
case with no such state. The third solution is a result obtained by
Newton \cite{Newtona}. Equation~(\ref{Levmodsstsol}) is the most
general statement of Levinson's theorem possible. The only input in
our derivation is the completeness relation~(\ref{compl}) and the
boundary conditions~(\ref{bcond}). Although we do not have an example
for the second solution in~(\ref{Levmodsstsol}) there is little doubt
that it can be realized.

Levinson's theorem~(\ref{Lev}) can be proved in a number of different
ways (see ref.~\cite{Jauch} and~\cite{Martin}), for instance). The
present proof has the advantage of being very general and extremely
simple and also allows us some additional insights.

\subsection{DIRAC CASE}

A further advantage of the above proof is that it can be taken over to
the case of the Dirac equation almost without change. In this section we
will first rederive the result of Ma and Ni \cite{MaNi},
which is the original correct statement of Levinson's theorem for Dirac
particles, and which is valid for the sum of positive and negative
energy phase shifts. Then we will prove a stronger statement of
Levinson's theorem, valid for positive and negative energies separately.
\par
The zero-energy resonance of the preceding chapter is now called
threshold resonance (its meaning is still the same: $k=0$). As in the
Schr\"odinger case we distinguish two cases: with and without threshold
resonance. However, there is now a slight complication - the threshold
resonance can have either a positive or negative energy, or there could
exist two resonances simultaneously, one with positive and the other
with negative energy. For the sake of simplicity we will consider only
two cases; the gap will be filled at the end of this section where we
consider the positive and negative energies separately. In the first
case any finite threshold solution (which does not necessarily exist) of
the Dirac equation~(\ref{raddir}) is a resonance and not a bound state. The
completeness relation in this case reads
\begin{eqnarray}\delta\left(r-r'\right)\delta_{ij}=\sum_{-m<\epsilon_\kappa<m}
u_{i\epsilon_\kappa\kappa}\left(r\right)\;u_{j\epsilon_\kappa\kappa}\left(r'
\right)\hspace{117pt}&&\nonumber\\+\int_0^\infty\frac{{\rm d}k}{2\pi}\left[
u_{i\varepsilon\kappa}\left(r\right)u_{j\varepsilon\kappa}\left(r'\right)+
u_{i,-\varepsilon\kappa}\left(r\right)u_{j,-\varepsilon\kappa}\left(r'\right)
\right]\,,&&\label{compld}\end{eqnarray}
where $i=1,2$, $j=1,2$, $u_{i\epsilon_\kappa\kappa}$ are bound state
solutions of~(\ref{raddir}), $\varepsilon=\sqrt{k^2+m^2}$ and where
the region of integration includes $k=0$, so that a term with a
delta-function $\delta\left(k\right)$ would contribute. The bound states are
normalized according to
\begin{equation}\int_0^\infty{\rm d}r\left[u_{1\epsilon_\kappa\kappa}^2\left(r
\right)+u_{2\epsilon_\kappa\kappa}^2\left(r\right)\right]=1\,.
\label{bstnormdintr}\end{equation}
In the second case any finite threshold solution (which does not
necessarily exist) is a bound state. The corresponding completeness
relation is
\begin{eqnarray}\delta\left(r-r'\right)\delta_{ij}=\sum_{-m\leq\epsilon_\kappa
\leq m}u_{i\epsilon_\kappa\kappa}\left(r\right)\;u_{j\epsilon_\kappa\kappa}
\left(r'\right)\hspace{117pt}&&\nonumber\\+\int_{0+}^\infty\frac{{\rm d}k}
{2\pi}\left[u_{i\varepsilon\kappa}\left(r\right)u_{j\varepsilon\kappa}\left(
r'\right)+u_{i,-\varepsilon\kappa}\left(r\right)u_{j,-\varepsilon\kappa}\left(
r'\right)\right]\,,&&\label{compldplus}\end{eqnarray}
where the region of integration does not include $k=0$, so that a term
with a delta-function $\delta\left(k\right)$ would not contribute.
The rest of the derivation is identical to the Schr\"odinger case, except
that equation~(\ref{diagndir})  has to be used instead
of~(\ref{diagn}). The result following from~(\ref{compldplus}) is
\begin{equation}\eta_{m\kappa}\left(0\right)+\eta_{-m,\kappa}\left(0\right)=
\left(N_\kappa^++N_\kappa^-\right)\pi\,,\label{Levd}\end{equation}
whereas the one following from~(\ref{compld}) is
\begin{equation}\eta_{m\kappa}\left(0\right)+\eta_{-m,\kappa}\left(0\right)=
\left(N_\kappa^++N_\kappa^-\right)\pi+\left(-1\right)^l\frac{\pi}{2}\left[
\sin^2\eta_{m\kappa}\left(0\right)+\sin^2\eta_{-m,\kappa}\left(0\right)\right]
\,,\label{Levmodd}\end{equation}
where $l=\vert\kappa\vert-1$ for $\kappa=-1,-2,\ldots$ and $l=\kappa$
for $\kappa=1,2,\ldots$ is the orbital angular momentum, $N_\kappa^+$
is the number of positive and $N_\kappa^-$ is the number of negative
energy bound states of~(\ref{raddir}). In deriving~(\ref{Levmodd}) we used
$\eta_{\infty\kappa}\left(\infty\right)=-\eta_{-\infty,\kappa}\left(
\infty\right)$ which follows from~(\ref{etakdlargea}).
Equation~(\ref{Levmodd}) is Levinson's theorem for Dirac particles
first obtained by Ma and Ni \cite{MaNi}. Above we derived
equations~(\ref{qdireqo}) and~(\ref{qdireqt}) which, if subjected to
the boundary conditions~(\ref{bcondo}) and~(\ref{bcondt}), are
equivalent to the Dirac equation~(\ref{raddir}). An inspection of these
equations at threshold ($k=0$) and for large $r$, shows that a threshold
resonance can occur only if $\kappa=\pm1$. Therefore~(\ref{Levd}) is valid
for all $\kappa=\pm2,\pm3,\ldots\,$ and for $\kappa=\pm1$ if there is
no threshold resonance, whereas~(\ref{Levmodd}) is valid for $\kappa=-1$
($l=0,\;\overline l=1$) and $\kappa=1$ ($l=1\;,\overline l=0$).

One may ask oneself whether~(\ref{Levd}) and~(\ref{Levmodd}) are
perhaps valid for positive and negative energies separately. In fact,
in some of the initial work (see refs.~\cite{Barth} and \cite{Ni})
this was claimed to be true but later was found incorrect
\cite{MaNi}. Yet, intuitively, by a number of reasons we
expect that an extension of~(\ref{Levd}) and~(\ref{Levmodd}), for positive and
negative energies separately, should be possible. We now show that
this is indeed the case. The basic observation which we need is the
fact that the set of Schr\"odinger-like equations~(\ref{wdireqo}) and
(\ref{wdireqt}) subject to the boundary conditions~(\ref{wbcondo})
and~(\ref{wbcondt}) is fully equivalent to the original Dirac
equation~(\ref{raddir}) subject to the boundary
conditions~(\ref{bcondo}) and~(\ref{bcondt}). Equations~(\ref{wdireqo})
and~(\ref{wdireqt}) are useful because they are not coupled and hence
the phase shift $\eta_{\epsilon\kappa}\left(k\right)$ can be computed using
any one of them, without reference to the other and for each of the
positive and negative energies $\epsilon$ separately. Certainly we
cannot apply Levinson's theorem to~(\ref{wdireqo}) or~(\ref{wdireqt}) directly
since the potential in these equations depends on the energy and hence
the completeness relation does not have the form~(\ref{compl}). However,
consider the following equations
\begin{equation}w_{kl}^+{''}-\left[\frac{l\left(l+1\right)}{r^2}-
\frac{\kappa}{r}\frac{V'}{2m-V}+\frac{1}{2}\frac{V''}{2m-V}+\frac{3}{4}
\left(\frac{V'}{2m-V}\right)^2-V^2+2mV-k^2\right]w_{kl}^+=0\,,
\label{wdireqp}\end{equation}
\begin{equation} w_{k\overline l}^-{''}-\left[\frac{\overline l\left(
\overline l+1\right)}{ r^2}-\frac{\kappa}{r}\frac{V'}{2m+V}-\frac{1}{2}
\frac{V''}{2m+V}+\frac{3}{4}\left(\frac{V'}{2m+V}\right)^2-V^2-2mV-k^2\right]
w_{k\overline l}^-=0\,,\label{wdireqm}\end{equation}
which are subject to the boundary conditions
\begin{equation} w_{kl}^+\left(0\right)=0\,,\quad w_{kl}^+\left(r\right)\;
\ar r\;2\,\sin\left(kr-\frac{\pi l}{2}+\eta_l^+\left(k\right)\right)\,,
\label{wbcondp}\end{equation}
\begin{equation} w_{k\overline l}^-\left(0\right)=0\,,\quad w_{k
\overline l}^-\left(r\right)\;\ar r\;2\,\sin\left(kr-\frac{\pi\overline l}{2}+
\eta_{\overline l}^-\left(k\right)\right)\,.\label{wbcondm}\end{equation}
At threshold ($k=0$)~(\ref{wdireqp}) and~(\ref{wbcondp}) coincide
with~(\ref{wdireqo}) and~(\ref{wbcondo}) for $\epsilon=m$, and
similarly~(\ref{wdireqm}) and~(\ref{wbcondm}) coincide
with~(\ref{wdireqt}) and~(\ref{wbcondt}) for $\epsilon=-m$. Moreover,
both sets of equations and boundary conditions are analytical near the
threshold. Therefore
\begin{equation}\eta_l^+\left(0\right)=\eta_{m\kappa}\left(0\right)\,,\quad
\eta_{\overline l}^-\left(0\right)=\eta_{-m,\kappa}\left(0\right)\,.
\label{etapetam}\end{equation}
Actually one would expect an ambiguity in both equations~(\ref{etapetam}),
each in terms of an additive integer multiple of $2\pi$. However, it
is easy to see that both integers (say $n_1$ and $n_2$) must be zero.
This follows from the fact that the simultaneous change  of $m$ to
$-m$ and $\kappa$ to $-\kappa$ is a symmetry operation, which implies
$n_1=n_2$.  Equation~(\ref{Levd}) (or~(\ref{Levmodd})), on the other
hand, implies $n_1=-n_2$, and hence both integers are zero.
Equations~(\ref{wdireqp}) and~(\ref{wdireqm}) are just usual
Schr\"odinger equations, so that we can apply Levinson's theorem
(\ref{Lev}) and~(\ref{Levmodsstsol}) and obtain
\begin{equation}\hspace{6.4pt}\eta_{m\kappa}\left(0\right)=\left\{\begin{array}
{ll}n_l^+\pi,&l=0,1,\ldots\\\left(n_0^++\frac{1}{4}\right)\pi,&l=0\\\left(n_0^+
+\frac{1}{2}\right)\pi,\quad&l=0\end{array}\,,\right.
\label{Levwdsolp}\end{equation}
\begin{equation}\eta_{-m\kappa}\left(0\right)=\left\{\begin{array}{ll}
n_{\overline l}^-\pi,&\overline l=0,1,\ldots\\\left(n_0^-+\frac{1}{4}\right)\pi
,&\overline l=0\\\left(n_0^-+\frac{1}{2}\right)\pi,\quad&\overline l=0
\end{array}\,,\right.\label{Levwdsolm}\end{equation}
where $n_l^+$ is the number of bound state solutions ($k^2<0$) of
(\ref{wdireqp}) and $n_{\overline l}^-$ is the number of bound state
solutions of (\ref{wdireqm}). In~(\ref{Levwdsolp})
and~(\ref{Levwdsolm}) the first case refers to a situation without a
threshold resonance and the other two cases to a situation with a
threshold resonance. Equations~(\ref{Levwdsolp}) and (\ref{Levwdsolm})
constitute Levinson's theorem for Dirac particles. As a consequence
of~(\ref{Levd}) and~(\ref{Levwdsolp}),~(\ref{Levwdsolm}) we have
\begin{equation}N_\kappa^++N_\kappa^-=n_l^++n_{\overline l}^-\,,
\label{Nnpm}\end{equation}
whereas in general $N_\kappa^+\neq n_l^+$ and
$N_\kappa^-\neq n_{\overline l}^-$. It is an interesting fact to
notice that~(\ref{wdireqp}) and (\ref{wdireqm}) do not correspond to
the usual expansion based on the Foldy-Wouthuysen scheme
(see~\cite{FW}). The above trick of "freezing" the energy of the
second order Dirac equation was used for a different purpose in
ref.~\cite{Grosse}.

\section{\hspace{-13.1pt}APPLICATION TO A NONPERTURBATIVE QED}

Consider the following set of equations
\begin{eqnarray}&&u'_{1\epsilon\kappa}+\frac{\kappa}{r}\,u_{1\epsilon\kappa}-
\left(m_0+\epsilon+\varphi\right)u_{2\epsilon\kappa}=0\nonumber\\
&&u'_{2\epsilon\kappa}-\frac{\kappa}{r}\,u_{2\epsilon\kappa}-\left(m_0-
\epsilon-\varphi\right)u_{1\epsilon\kappa}=0\,,\label{radQEDdir}\end{eqnarray}
\begin{equation}\varphi''+\frac{2}{r}\varphi'=\frac{e_0^2}{4\pi}\;\frac{1}
{r^2}\sum_{\kappa=1}^\infty\kappa\left(\varrho_{\kappa}+\varrho_{-\kappa}
\right)\,,\label{radQEDmaxw}\end{equation}
\begin{eqnarray}&&\delta\left(r-r'\right)\delta_{ij}=\sum_{0\leq
\varepsilon_\kappa\leq m_0}u_{i\varepsilon_\kappa\kappa}\left(r\right)\;
u_{j\varepsilon_\kappa\kappa}\left(r'\right)+\sum_{0<\varepsilon_\kappa\leq
m_0}u_{i,-\varepsilon_\kappa\kappa}\left(r\right)\;u_{j,-\varepsilon_\kappa
\kappa}\left(r'\right)\nonumber\\ &&\hspace{75pt}+\int_{0+}^\infty
\frac{{\rm d}k}{2\pi}\left[u_{i\varepsilon\kappa}\left(r\right)u_{jE\kappa}
\left(r'\right)+u_{i,-\varepsilon\kappa}\left(r\right)u_{j,-\varepsilon\kappa}
\left(r'\right)\right]\,,\label{radQEDcompl}\end{eqnarray}
where $i=1,2$, $j=1,2$, $\kappa=\pm1,\pm2,\ldots\,$,
$u_{i,\pm \varepsilon_\kappa\kappa}$ are bound state and
$u_{i,\pm \varepsilon\kappa}$ are scattering state solutions
of~(\ref{radQEDdir}), $\epsilon=\pm\varepsilon_\kappa$,
$0\leq\varepsilon_\kappa\leq m_0$, are bound state and
$\epsilon=\pm\varepsilon$, $\varepsilon\equiv\sqrt{m_0^2+k^2}$, are
scattering state energies and
\begin{eqnarray}&&\varrho_\kappa=\sum_{0\leq \varepsilon_\kappa\leq m_0}\left(
u_{1\varepsilon_\kappa\kappa}^2+u_{2\varepsilon_\kappa\kappa}^2\right)-\sum_{
0<\varepsilon_\kappa\leq m_0}\left(u_{1,-\varepsilon_\kappa\kappa}^2+u_{2,
-\varepsilon_\kappa\kappa}^2\right)\nonumber\\ &&\hspace{25pt}+\int_{0+}^\infty
\frac{{\rm d}k}{2\pi}\left(u_{1\varepsilon\kappa}^2+u_{2\varepsilon\kappa}^2-
u_{1,-\varepsilon\kappa}^2 -u_{2,-\varepsilon\kappa}^2\right)\,.
\label{radQEDrho}\end{eqnarray}
For the sake of simplicity we assume that there are no threshold
resonances and hence in~(\ref{radQEDcompl}) and~(\ref{radQEDrho}) the region of
integration excludes $k=0$. The boundary conditions are
\begin{equation}\left|\varphi\left(0\right)\right|<\infty\,,\qquad\varphi\left(
\infty\right)=0\label{bcondmaxw}\end{equation}
for the field $\varphi$, and the usual boundary conditions
\begin{equation}u_{i,\pm\varepsilon_\kappa\kappa}\left(0\right)=0\,,\qquad
u_{i,\pm\varepsilon_\kappa\kappa}\left(\infty\right)\;\ar r\;c_{i,\pm
\varepsilon_\kappa\kappa}\,{\rm e}^{-\gamma_\kappa r}\,,
\label{bcondbst}\end{equation}
hold for the bound states, where $c_{i,\pm\varepsilon_\kappa\kappa}$ are real
constants and $\gamma_\kappa=\sqrt{m_0^2-\varepsilon_\kappa^2}\geq0$, and
\begin{equation}u_{1\epsilon\kappa}\left(0\right)=0\,,\quad
u_{1\epsilon\kappa}\left(r\right)\;\ar r\;\,\sqrt{\frac{\epsilon+m_0}
{2\epsilon}}\,2\,\sin\left(kr-\frac{\pi l}{2}+\eta_{\epsilon\kappa}\left(k
\right)\right)\label{QEDbcondo}\end{equation}
\begin{equation}u_{2\epsilon\kappa}\left(0\right)=0\,,\quad
u_{2\epsilon\kappa}\left(r\right)\;\ar r\;\,\sqrt{\frac{\epsilon+m_0}
{2\epsilon}}\,\frac{\kappa_0k}{\epsilon+m_0}\,2\,\sin\left(kr-\frac{\pi
\overline l}{2}+\eta_{\epsilon\kappa}\left(k\right)\right)
\label{QEDbcondt}\end{equation}
for the scattering states, where
$l=\vert\kappa\vert-\left(1-\kappa_0\right)/2$, $\overline l=l-\kappa_0$
and $\kappa_0=\kappa/\vert\kappa\vert$. The bound states are
normalized according to
\begin{equation}\int_0^\infty{\rm d}r\left[u_{1,\pm\varepsilon_\kappa
\kappa}^2\left(r\right)+u_{2,\pm\varepsilon_\kappa\kappa}^2\left(r\right)
\right)=1\,.\label{bstnormd}\end{equation}
Notice that~(\ref{radQEDdir}-\ref{radQEDcompl}) is a closed system of
equations. Also notice that any solution of this system constitutes a
complete orthonormal set of functions and, therefore, Pauli's
exclusion principle is built in.
Equations~(\ref{radQEDdir}-\ref{radQEDcompl}) arise as a spherically
symmetric special case of more general equations in a recently
proposed generalized interaction picture of quantum electrodynamics
(QED). This generalized interaction picture is useful, since it allows
for a new nonperturbative approach to QED. The derivation of these
equations from QED and more details on their properties will be given
in a forthcoming paper \cite{polQED}. Here we want to apply the
results of the preceding sections to the investigation of
self-consistency of equations~(\ref{radQEDdir}-\ref{radQEDcompl}). We shall
prove in this section that for any solution of
(\ref{radQEDdir}-\ref{radQEDcompl}) with a finite number of bound states
the total charge vanishes: $Q_0=0$, where the charge density is defined
by the right-hand side of~(\ref{radQEDmaxw}) (in units of $-e_0^2$).
Furthermore, the coupling constant $e_0^2/4\pi$ is not a free parameter
but rather must have a numerical value for which $\int_0^\infty{\rm d}r\;
\varphi\left(r\right)=0$.\par
Firstly we notice that there exists at least one solution. In fact,
if we assume that there are no bound states then
\begin{equation}u_{1\epsilon\kappa}= v_{1\epsilon\kappa}\,,\qquad
u_{2\epsilon\kappa}= v_{2\epsilon\kappa}\label{QEDzNsol}\end{equation}
is a solution of~(\ref{radQEDdir}-\ref{radQEDcompl}), where
\begin{equation}v_{1\epsilon\kappa}\left(r\right)=\sqrt{\frac{\epsilon+m_0}
{2\epsilon}}\,2kr\,j_l\left(kr\right)\,,\quad v_{2\epsilon\kappa}\left(r
\right)=\sqrt{\frac{\epsilon+m_0}{2\epsilon}}\,\frac{\kappa_0k}{\epsilon+m_0}\,
2kr\,j_{\overline l}\left(kr\right)\label{freewf}\end{equation}
and $ j_l\left(kr\right)$,  $j_{\overline l}\left(kr\right)$  are the
spherical Bessel functions. In general, however, a solution
of~(\ref{radQEDdir}-\ref{radQEDcompl}) will contain a number of positive
and a number of negative bound states. While the possibility of an
infinite number of bound states cannot be ruled out on simple grounds,
we want to examine the consistency conditions for a finite number of
bound states.

If we multiply~(\ref{radQEDmaxw}) by $r^2$ and integrate the resulting
equation we obtain
\begin{equation}\varphi'\left(r\right)=-\frac{e_0^2}{4\pi}\,\frac{Q_0}{r^2}-
\frac{e_0^2}{4\pi}\,\frac{1}{r^2}\sum_{\kappa=1}^\infty\kappa\int_r^\infty
{\rm d}r'\left[\varrho_{\kappa}\left(r'\right)+\varrho_{-\kappa}\left(r'
\right)\right]\,,\label{phiprime}\end{equation}
where
\begin{equation}Q_0=-\sum_{\kappa=1}^\infty\kappa\int_0^\infty{\rm d}r'\left[
\varrho_{\kappa}\left(r'\right)+\varrho_{-\kappa}\left(r'\right)\right]\,.
\label{Qnot}\end{equation}
At large $r$ we have
\begin{eqnarray}\varrho_\kappa\left(r\right)\;\ar r\;\sum_{0\leq
\varepsilon_\kappa\leq m_0}\left(c_{1\varepsilon_\kappa\kappa}^2+
c_{2\varepsilon_\kappa\kappa}^2\right)\,{\rm e}^{-\gamma_\kappa r}-
\sum_{0<\varepsilon_\kappa\leq m_0}\left(c_{1,-\varepsilon_\kappa\kappa}^2+
c_{2,-\varepsilon_\kappa\kappa}^2\right)\,{\rm e}^{-\gamma_\kappa r}
\hspace{30pt}&&\nonumber\\-\left(-1\right)^l4m\int_{0+}^\infty\frac{{\rm d}k}
{2\pi}\,\frac{1}{\varepsilon}\,\cos\left[\eta_{\varepsilon\kappa}\left(k\right)
-\eta_{-\varepsilon\kappa}\left(k\right)\right]\cos\left[2kr+\eta_{\varepsilon
\kappa}\left(k\right)+\eta_{-\varepsilon\kappa}\left(k\right)\right]&&
\label{rhorlarge}\end{eqnarray}
and hence
\begin{eqnarray}\int_r^\infty{\rm d}r'\varrho_\kappa\left(r'\right)\;\ar r\;
\sum_{0\leq\varepsilon_\kappa\leq m_0}\frac{c_{1\varepsilon_\kappa\kappa}^2+
c_{2\varepsilon_\kappa\kappa}^2}{\gamma_\kappa}\,{\rm e}^{-\gamma_\kappa r}-
\sum_{0<\varepsilon_\kappa\leq m_0}\frac{c_{1,-\varepsilon_\kappa\kappa}^2+
c_{2,-\varepsilon_\kappa\kappa}^2}{\gamma_\kappa}\,{\rm e}^{-\gamma_\kappa r}
\hspace{25pt}&&\nonumber\\+\left(-1\right)^l2m\int_{0+}^\infty\frac{{\rm d}k}
{2\pi}\,\frac{1}{\varepsilon k}\,\cos\left[\eta_{\varepsilon\kappa}\left(k
\right)-\eta_{-\varepsilon\kappa}\left(k\right)\right]\sin\left[2kr+
\eta_{\varepsilon\kappa}\left(k\right)+\eta_{-\varepsilon\kappa}\left(k\right)
\right].&&\label{intrhorlarge}\end{eqnarray}
It now follows from~(\ref{phiprime}) and~(\ref{intrhorlarge}) that
\begin{equation}\varphi\left(r\right)\;\ar r\;\frac{e_0^2}{4\pi}\,\frac{Q_0}
{r}+O\left({\rm e}^{-ar}\right)\,,\quad a>0\,.\label{phirlarge}\end{equation}
Now, if $Q_0>0$ then $\varphi$ acts as an attractive potential for the
positive energy states and as a repulsive potential for the negative
energy states. Similarly, if $Q_0<0$ then $\varphi$ acts as an
attractive potential for the negative energy states and as a repulsive
potential for the positive energy states. Consequently, in both cases
$\varphi$ supports an infinite number of bound states, like any
attractive potential, which vanishes at infinity not faster than
$1/r^2$. Thus $Q_0>0$ and $Q_0<0$ are inconsistent with a finite number
of bound states. On the other hand, if $Q_0=0$ then there will be at
most a finite number of bound states, since $\varphi$ acts as a
short range potential with nonvanishing repulsive and attractive parts.
The repulsive part is due to the electrostatic self-interaction of bound
states, whilst the attractive part is provided by the scattering
states, which fully compensate the repulsive part. Thus the consistency
condition for the existence of a finite number of bound states is
\begin{equation}Q_0=0\,.\label{QEDcons}\end{equation}
\par
This consistency condition implies a restriction on $\varphi$.
Combining~(\ref{Qnot}) and~(\ref{radQEDrho}), we obtain
\begin{equation}Q_0=-\sum_{\kappa=-\infty}^\infty\vert\kappa\vert\left\{
N_\kappa^+-N_\kappa^-+\int_{0+}^\infty\frac{{\rm d}k}{2\pi}\int_0^\infty
{\rm d}r\left[u_{1\varepsilon\kappa}^2\left(r\right)+u_{2\varepsilon\kappa}^2
\left(r\right)-u_{1,-\varepsilon\kappa}^2\left(r\right)-u_{2,-\varepsilon
\kappa}^2\left(r\right)\right]\right\}\,.\label{Qnota}\end{equation}
where $N_\kappa^+$ and $N_\kappa^-$ are the number of positive and
negative energy bound states of~(\ref{radQEDdir}) respectively. Substituting
(\ref{diagndira}), we get
\begin{equation}Q_0=-\sum_{\kappa=-\infty}^\infty\vert\kappa\vert\left\{
N_\kappa^+-N_\kappa^--\left[\eta_{m\kappa}\left(0\right)-\eta_{-m\kappa}\left(
0\right)\right]\frac{1}{\pi}+\left[\eta_{\infty\kappa}\left(\infty\right)-
\eta_{-\infty\kappa}\left(\infty\right)\right]\frac{1}{\pi}\right\}\,.
\label{Qnotb}\end{equation}
Since the boundary conditions for $\varphi$ prohibit a singularity at
the origin, we can use~(\ref{etakdlarge}) for
$\eta_{\pm\infty}\left(\infty\right)$ and obtain
\begin{equation}Q_0=-\sum_{\kappa=-\infty}^\infty\vert\kappa\vert\left\{
N_\kappa^+-N_\kappa^--n_l^++n_{\overline l}^-+\frac{2}{\pi}\int_0^\infty{\rm d}
r\;\varphi\left(r\right)\right\}\,,\label{Qnotc}\end{equation}
where we used Levinson's theorem~(\ref{Levwdsolm}), and where $n_l^+$ and
$n_{\overline l}^-$ are the number of bound state solutions of~(\ref{wdireqp})
and (\ref{wdireqm}) with $V=-\,\varphi$ and $m=m_0$ respectively. Since the
last term in~(\ref{Qnotc}) is independent of $\kappa$ it must vanish,
otherwise $Q_0$ would become infinite. Hence we conclude that
\begin{equation}\int_0^\infty{\rm d}r\;\varphi\left(r\right)=0\label{QEDconsa}
\end{equation}
must hold for any solution of~(\ref{radQEDdir}-\ref{radQEDcompl}) having a
finite number of bound states. Notice that if all of the bound states
have either positive or negative energies, then~(\ref{QEDconsa})
implies~(\ref{QEDcons}), which follows from (\ref{Nnpm}). Also notice
that $\int_0^\infty{\rm d}r\,\varphi\left(r\right)$ is a dimensionless
quantity and therefore does not depend on the bare mass parameter
$m_0$ and is a function of the coupling constant $e_0^2/4\pi$ only.
Hence, due to~(\ref{QEDconsa}), the coupling constant $e_0^2/4\pi$
becomes an eigenvalue of~(\ref{radQEDdir}-\ref{radQEDcompl}), i.e. we
expect that (\ref{QEDconsa}) holds only if $e_0^2/4\pi$ acquires one
or more specific numerical values. To illuminate this point we
consider the following iteration procedure. As the initial step we
make a reasonable guess of $\varphi$, which satisfies~(\ref{QEDconsa})
and which we denote by $\varphi_0$. Then we solve the Dirac
equation~(\ref{radQEDdir}) with $\varphi=\varphi_0$ and obtain a
complete set of bound and scattering states. At this stage
$e_0^2/4\pi$ is a free parameter. Now we fix it by requiring that with
this complete set the solution of~(\ref{radQEDmaxw}), which we denote
by $\varphi_1$, satisfies~(\ref{QEDconsa}). If there is no value of
$e_0^2/4\pi$ for which~(\ref{QEDconsa}) is satisfied, then we start
again trying to make a better choice of $\varphi_0$. If, however,
there is such a value of $e_0^2/4\pi$, then we have completed the first
iteration and can start the second one, now with $\varphi_1$ instead
of $\varphi_0$. Assume that this iteration procedure is convergent,
then it is clear that, as a result of~(\ref{QEDconsa}), the coupling
constant will emerge, fixed at one or several numerical values.

\section*{Acknowledgments}

The author wishes to thank Y. Eisenberg for valuable discussions and
suggestions. Thanks are also due to G. Eden for reading the paper
and making useful comments.


\begin{thebibliography}{99}
\bibitem[\ *]{Email}{Present address: Theoretische Physik,
ETH-H\"onggerberg, CH-8093 Z\"urich, Switzerland. E-mail:
poli@itp.ethz.ch}
\bibitem{Levinson}{N. Levinson, On the uniqueness of the potential in a
Schr\"odinger equation for a given asymptotic phase. Kgl. Danske
Videnskab. Selskab, Mat.-fys. Medd., {\bf 25}(9) (1949).}
\bibitem{Spruchprl}{Z. R. Iwinski, Leonard Rosenberg, and Larry
Spruch, Phys. Rev. Lett., 1602 (1985).}
\bibitem{Spruchpr}{Z. R. Iwinski, Leonard Rosenberg, and Larry
Spruch, Phys. Rev. A {\bf33}, 946 (1986).}
\bibitem{Niemi}{A. J. Niemi and G. W. Semenoff, Phys. Rev. D {\bf32},
471 (1985).}
\bibitem{Blank}{R. Blankenbecler and D. Boyanovsky, Physica 18D, 367
(1986).}
\bibitem{Friedel}{J. Friedel, Nuovo Cim. Suppl., vol. 7, serie 10, 287
(1958).}
\bibitem{Newtona}{R. G. Newton, Analytic Properties of Radial Wave
Functions. J. Math. Phys. {\bf 1}, 319 (1960).}
\bibitem{Ma}{Z. Q. Ma, Proof of the Levinson theorem by the Sturm-Liouville
theorem. J. Math. Phys. {\bf26}, 1995 (1985). See also J. Phys.
A: Math. Gen. {\bf21}, 2085 (1988).}
\bibitem{Ni}{G.-J. Ni, The Levinson theorem and its generalization
in relativistic quantum mechanics. Phys. Energ. Fortis {\&} Phys. Nucl.
(China), vol. 3, no. 4, p. 432-49 (1979). See also ref.~\cite{Ma}.}
\bibitem{MaNi}{Z. Q. Ma, G.-J. Ni, Levinson theorem for Dirac particles.
Phys. Rev. D {\bf31}, 1482 (1985). See also Phys. Rev. D
{\bf32}, 2203 (1985), Phys. Rev. D {\bf32}, 2213 (1985).}
\bibitem{Newton}{R. G. Newton, {\it Scattering Theory of Waves and
Particles} (second edition, Springer-Verlag, New York, 1982).}
\bibitem{Parzen}{G. Parzen, On the Scattering Theory of the Dirac Equation.
Phys. Rev. {\bf80}, 261 (1950).}
\bibitem{polQED}{N. Poliatzky, Generalized Interaction Picture of Quantum
Electrodynamics. Preprint WIS-92/104/Dec-Ph.}
\bibitem{Baz}{A. I. Baz', Ya. B. Zeldovich, A. M. Perelomov, {\it Scattering,
Reactions and Decay in Nonrelativistic Quantum Mechanics} (Jerusalem,
1969), p. 61.}
\bibitem{Lued}{G. L\"uders, Zum Zusammenhang zwischen S-Matrix und
Normierungsintegralen in der Quantenmechanik. Z. f. Naturforsch.,
{\bf 10a}, 581 (1955).}
\bibitem{Taylor}{J. R. Taylor, {\it Scattering Theory} (John
Wiley~\&~Sons, New York, 1972).}
\bibitem{Jauch}{J. M. Jauch, On the relation between scattering phase
and bound states. Helv. Phys. Acta, {\bf30}, 143 (1957).}
\bibitem{Martin}{A. Martin, On the Validity of Levinson's Theorem for
Non-Local Interactions. Nuovo Cim. {\bf7}, 607 (1958).}
\bibitem{Barth}{M.-C. Barth\'el\'emy, Ann. Inst. Henri Poincar\'e, {\bf7},
115 (1967).}
\bibitem{FW}{J. D. Bjorken, S. D. Drell, {\it Relativistic Quantum
Mechanics} (McGraw-Hill, New York, 1964).}
\bibitem{Grosse}{H. Grosse, A. Martin, J. Stubbe, Order of energy levels
in relativistic one-electron models. Preprint CERN-TH.6343/91.}
\end{thebibliography}
\end{document}